\documentclass[sigconf, authorversion]{acmart}
\usepackage[utf8]{inputenc}
\usepackage{booktabs}
\usepackage{multirow}
\usepackage{microtype}
\usepackage{wasysym}
\usepackage{color,soul}
\usepackage{tabularx}
\usepackage{hyperref}

\copyrightyear{2023}
\acmYear{2023}
\setcopyright{acmlicensed}\acmConference[CHI '23]{Proceedings of the 2023 CHI Conference on Human Factors in Computing Systems}{April 23--28, 2023}{Hamburg, Germany}
\acmBooktitle{Proceedings of the 2023 CHI Conference on Human Factors in Computing Systems (CHI '23), April 23--28, 2023, Hamburg, Germany}
\acmPrice{15.00}
\acmDOI{10.1145/3544548.3580967}
\acmISBN{978-1-4503-9421-5/23/04}

\begin{document}

\author{William Seymour}
\email{william.1.seymour@kcl.ac.uk}
\orcid{0000-0002-0256-6740}
\affiliation{%
  \institution{King's College London}
  \streetaddress{Bush House, 30 Aldwych}
  \city{London}
  \country{UK}
  \postcode{WC2B 4BG}
}

\author{Mark Cot\'{e}}
\email{mark.cote@kcl.ac.uk}
\affiliation{%
  \institution{King's College London}
  \streetaddress{Chesham Building, Strand}
  \city{London}
  \country{UK}
  \postcode{WC2R 2LS}
}

\author{Jose Such}
\email{jose.such@kcl.ac.uk}
\affiliation{%
  \institution{King's College London}
  \streetaddress{Bush House, 30 Aldwych}
  \city{London}
  \country{UK}
  \postcode{WC2B 4BG}
}

\keywords{Voice Assistants, Consent, Verbal Consent, Informed Consent, GDPR, Alexa, Conversational User Interfaces, Permissions}
\title{Legal Obligation and Ethical Best Practice: Towards Meaningful Verbal Consent for Voice Assistants}

\ccsdesc[500]{Security and privacy~Social aspects of security and privacy}
\ccsdesc[300]{Security and privacy~Usability in security and privacy}
\ccsdesc[500]{Human-centered computing~Natural language interfaces}
\ccsdesc[300]{Applied computing~Law}

\begin{abstract}
To improve user experience, Alexa now allows users to consent to data sharing via voice rather than directing them to the companion smartphone app. While verbal consent mechanisms for voice assistants (VAs) can increase usability, they can also undermine principles core to informed consent. We conducted a Delphi study with experts from academia, industry, and the public sector on requirements for verbal consent in VAs. Candidate requirements were drawn from the literature, regulations, and research ethics guidelines that participants rated based on their relevance to the consent process, actionability by platforms, and usability by end-users, discussing their reasoning as the study progressed. We highlight key areas of (dis)agreement between experts, deriving recommendations for regulators, skill developers, and VA platforms towards crafting meaningful verbal consent mechanisms. Key themes include approaching permissions according to the user’s ability to opt-out, minimising consent decisions, and ensuring platforms follow established consent principles.
\end{abstract}

\maketitle

\section{Introduction}\label{sec:introduction}
The last decade has seen the widespread introduction of voice assistants (VAs) into domestic life in many parts of the world. Offering novelty and convenience, VAs have transformed the home computing landscape and have been positioned by vendors as the centre of the smart home as a hub for other apps and gadgets: research shows they are most commonly used to play music, search for information, and control other IoT devices~\cite{ammari2019music}. In this way their usage extends that of the smartphone where app stores allow for the use of a wide variety of third-party software, and many popular smartphone apps are also available as skills/actions, including smart device companion apps. However, VAs do not simply offer access to traditional means of computing via a new interaction modality, their design and interfaces also represent a shift in people's underlying relationship with the technology that they use~\cite{nass1994computers, seymour2021exploring}.

A key component of VAs and other smart platforms is sharing data between/about users and third parties in order to enable extra functionality (e.g. a weather forecast that automatically accesses the user's location). However, developing mechanisms to facilitate this has been a persistent problem~\cite{felt2012android, utz2019uninformed}; solutions need to meaningfully let users choose what data to share with skill developers whilst balancing usability and legal obligations around data protection. Since their inception, popular VAs have used the same model as smartphone app stores, where a list of permissions is presented graphically on first use or at runtime with options to accept or decline.\footnote{\url{https://developer.amazon.com/en-US/docs/alexa/custom-skills/configure-permissions-for-customer-information-in-your-skill.html} \\ (accessed 18/11/2022)} For users this means switching to the assistant's companion smartphone app in order to use skills that require permissions. While this approach works well for smartphone apps where people are already using their phones, it is much less streamlined for VAs where users report the hands-free convenience of voice interaction as a key reason for use~\cite{ammari2019music}.

To address this, VA platforms are beginning to move these consent decisions into conversations: with Alexa's `Voice Forward Consent' feature (VFC),\footnote{\url{https://developer.amazon.com/en-US/docs/alexa/custom-skills/use-voice-forward-consent.html} (accessed 18/11/2022)} the VA reads out a list of requested permissions to which the user responds `I approve' or `no'. While potentially more usable than app-based alternatives, the switch to speech combined with other aspects of VA conversational design fundamentally changes the nature of the consent-granting process. Examples include the lack of audible distinction between platform and developer originated speech, and the pressure caused by timeout periods built into VA conversations~\cite{seymour2022can}. A key motivation for this work, therefore, is promoting the principles that underpin informed consent in the mechanisms that manage data collection and permissions within voice assistant ecosystems. Different perspectives on consent emphasise a variety of requirements for the verbal consent process, legal or otherwise, and in this work we explore potential such requirements drawn from the academic literature, data protection regulation, and research ethics guidelines.

To evaluate these requirements we present the results of a Delphi study that facilitated a discussion between subject experts from academia, industry, and the regulatory and policy sector. In so doing, we draw out the most important requirements for VA verbal consent along axes of relevance, actionability, and usability, and highlight areas of disagreement where the debate is ongoing. These insights are used to lay out recommendations for current and future implementations of verbal consent in VAs and other conversational interfaces. While we frequently use Alexa as an example of verbal consent in voice assistants as it has the most developed and documented verbal consent mechanism, our findings apply more generally. With conversational interfaces becoming increasingly prevalent in daily life and embedded in phones, TVs, headphones, and other devices we believe that considering these questions \textit{now} is of the utmost importance. The capabilities of these interfaces, both in terms of functionality and the amount of personal data they are able to access, will only increase with time. As such, we need to align how verbal consent is managed and perceived before bad practices are embedded in conversational interfaces that will later prove difficult or impossible to change (c.f. the lasting effects of the EU ePrivacy directive/`Cookie Law').

To this end, this paper answers the following research questions:
\begin{enumerate}
    \item[RQ1.] What consent requirements from regulation, the literature, and research ethics do experts agree are the most relevant, actionable, and usable for verbal consent in VAs? 
    \item[RQ2.] What are the areas of disagreement between experts around these requirements?
    \item[RQ3.] What changes should be made to current VA verbal consent processes to better align them with the requirements and principles that experts consider important?
\end{enumerate}

By answering these research questions, we  make the following contributions:
\begin{itemize}
    \item We show how consent plays a dual role in VAs and similar systems as a legal obligation and ethical best practice.
    \item We identify seven highly relevant, actionable, and usable requirements that should be implemented in VA verbal consent mechanisms.
    \item We derive six longer-term recommendations from expert discussion towards meaningful verbal consent in voice assistants.
\end{itemize}

In answering these questions it is important to note that our perspective on this issue is European, and we assume throughout the paper that the General Data Protection Regulation (GDPR) applies. While the direct applicability of the resultant analysis to the rest of the world is limited, many countries around the world have since enacted data protection laws based on or inspired by the GDPR. At the time of writing the European Commission recognises equivalent regulations in Andorra, Argentina, Canada, Faroe Islands, Guernsey, Israel, Isle of Man, Japan, Jersey, New Zealand, Republic of Korea, Switzerland, the United Kingdom, and Uruguay as providing adequate protection to allow data flows without additional safeguards.\footnote{\url{https://ec.europa.eu/info/law/law-topic/data-protection/international-dimension-data-protection/adequacy-decisions_en} (accessed 18/11/2022)}

\section{Background and Related Work}
\subsection{Voice Assistants, Skills, and Voice-Forward Consent}
Broadly speaking, voice assistants listen to nearby conversations trying to detect the `wake word' that signifies the beginning of a command. Once this has been detected they record the speech that follows and send it to a cloud service for transcription. This transcript is parsed to determine which feature (`skill') the user is asking for from a selection of thousands made by first- and third-party developers~\cite{edu2020smart}. A text response from the skill is then transformed into speech and sent back to the voice assistant device. Some devices can also show graphical prompts on connected screens called `cards'.

The most common model for these skills is based around \textit{intents}, \textit{utterances}, and \textit{slots}.\footnote{\url{https://developer.amazon.com/en-US/docs/alexa/custom-skills/create-intents-utterances-and-slots.html} (accessed 18/11/2022)} Intents represent ways of capturing actions users want to take within an interaction with the assistant: this could be a launch intent that runs when a skill opens (`Alexa, open Ride Hailer'), or an action that users take within a skill like starting a timer. Each intent has one or more utterances that reflect what the user must say in order to trigger them (e.g., ``open Ride Hailer''). Finally, intents may want to capture additional information using slots (e.g., ``what's the weather [in Paris]''). Together, these three elements represent the `front end' of a skill registered in a service called `Alexa Skills Kit' (ASK) and provide enough information for platforms to facilitate conversational interactions. The actual code for skills is self-hosted by developers (although many choose to utilise first-party service like AWS Lambda and Google Firebase).

Alexa permissions are declared in ASK and skills can check whether permissions have been granted to them when they run. Typically skills request permissions by generating a `consent card' and sending it to the user's device where it is read out and may appear on screen. The user is then prompted to enable the corresponding permissions in the Alexa smartphone app. In contrast to this hands-on way of asking for and granting permissions, Alexa now offers an alternative verbal consent mechanism called `Voice-Forward Consent'. When using VFC: (1) the third-party skill verbally justifies why it needs access to personal data; (2) the VA reads out the permissions requested by the third-party skill and signposts valid responses to the consent decision; (3) the user states their decision. A sample conversation snippet is shown in Table~\ref{tab:sampleVFC}.

\begin{table}
    \centering
    \caption{Sample VFC conversation from \url{https://developer.amazon.com/en-US/docs/alexa/custom-skills/use-voice-forward-consent.html}.}
    \begin{tabularx}{\columnwidth}{l|X}
    \textbf{User} & \textbf{Alexa, open Ride Hailer.}\\
    Alexa & Welcome to Ride Hailer. Where would you like to go?\\
    \textbf{User} & \textbf{The Space Needle.}\\
    Alexa & Sure. I need access to your name, current location, and mobile number so that I can find a ride for you.\\
    Alexa (OS) & Do you give Ride Hailer permission to access your name, current location, and mobile number? You can say `I approve' or `no'.\\
    \textbf{User} & \textbf{I approve.}\\
    Alexa & Thank you. A ride to the Space Needle from your current location will cost fifteen dollars, and the driver can pick you up in ten minutes. Do you want me to book it?\\
    \textbf{User} & \textbf{Yes.}\\
    Alexa & Great. Your driver will arrive in ten minutes.
    \end{tabularx}
    \label{tab:sampleVFC}
\end{table}

\subsection{Privacy and Security Concerns with Voice Assistants}
Prior work has identified a number of ethical concerns with voice assistants, primarily around privacy and social interactions. It is generally accepted that people have vague and/or incorrect mental models of how voice assistants work, including awareness of privacy controls~\cite{abdi2019more,abdi2021privacy, ammari2019music, huang2020amazon}. A recurring example of this is the belief by users that VAs are actively listening all of the time, and events such as false activations during normal conversations (and the resulting recordings) often cast further doubt that vendors are acting in good faith~\cite{meng2021owning}. Furthermore, users often believe that all skills for the Alexa platform are produced by Amazon~\cite{sabir2022hey}; if users do not realise that they are interacting with software developed by third parties then they cannot give informed consent to share data with them. One of the reasons given by people for continuing to use VAs despite having concerns is their trust in external protections, particularly privacy regulations~\cite{meng2021owning}.

There are also concerns about how the social nature of speech might allow VAs to manipulate people. Early research on the `computers are social actors' paradigm showed that people apply social stereotypes to computers, respond to them as if they were people, and reciprocate information sharing even when they know that they are interacting with a machine~\cite{nass1994computers, nass2000machines}. Work on anthropomorphism in voice assistants also suggests that using speech as an interaction modality can be pleasing and correlates with trust~\cite{seymour2021exploring, lopatovska2019talk}. Interest in more proactive assistants (i.e., that can take actions that do not immediately follow user requests) has raised questions around how information sharing and permissions would need to be adapted, especially around sensitive topics like finances or health~\cite{malkin2022runtime}.

Work looking at voice assistant platforms suggests that there are also problems with the way that skills are certified and moderated. Current work on the certification process suggests that initial checks miss much of the skill conversation tree, and that skills can be crafted to minimise testing coverage by human and automated checks~\cite{wang2021demystifying}. Similar work examining the efficacy of skill certification showed that policy-violating skills were approved for public use in over 60\% cases across the Alexa and Google Assistant skill stores~\cite{cheng2020dangerous}. Finally, others have studied data collection by skills, highlighting the prevalence of third-party software that collects personal data without using the mandatory permissions API and/or having privacy policies that are broken or deficient~\cite{edu2021skillvet, edu2022measuring, guo2020skillexplorer}. These works show that the number of available skills collecting personal data with broken or problematic privacy policies has been reducing year-on-year but still stands at over a third (36\% in 2021).

\subsection{Consent in Human-Computer Interaction and Privacy Regulation}\label{sec:consent-background}
There has been a growing strand of work relating feminist theories of consent centred around interpersonal contexts to challenges in Human-Computer Interaction (HCI) and Ubiquitous computing (Ubicomp). At a basic level, the fundamental concepts remain the same, with affirmative consent being a social process that is voluntary, informed, revertible, specific, and unburdensome~\cite{im2021yes}. Others have drawn directly on work around sexual consent, highlighting the importance of being able to easily explicate boundaries and withdraw consent during interactions with anthropomorphised devices like voice assistants~\cite{strengers2021what}. This work also suggests the possibility of developing ongoing dialogues around human-device consent as circumstances change over time. Through expert interviews on consent for Ubicomp, \citeauthor{luger2013informed} show a divide between those for whom consent is similar to a contract between users and device manufacturers, and those for whom it was more associated with rights and freedoms, enabling selective access to the self~\cite{luger2013informed}. In a research context, this more social reading of consent was seen to allow the person seeking consent to show respect to the person being asked and for that person to feel comfortable with the process they were being asked to consent to, and aligns with prior work showing that people readily perceive conversational agents as having personality aspects such as `respectful'~\cite{volkel2020developing}. This contrasts with the more contractual view where consent is a transfer of power, often to the already powerful~\cite{luger2013informed}.

Many of the principles underpinning consent in the literature are reflected in the GDPR, which defines consent as a ``freely given, specific, informed and unambiguous indication of the data subject’s wishes'' (Art. 4.11). Echoing the principles described by~\citeauthor{im2021yes}, in many cases the GDPR even uses the same language to describe consent~\cite{GDPR16}. For example:

\begin{itemize}
    \item \textbf{Voluntary}: e.g., ``In order to ensure that consent is freely given, consent should not provide a valid legal ground for the processing of personal data in a specific case where there is a clear imbalance between the data subject and the controller'' (Recital 43)
    \item \textbf{Informed}: ``For consent to be informed, the data subject should be aware at least of the identity of the controller and the purposes of the processing for which the personal data are intended'' (Recital 42)
    \item \textbf{Specific}: ``Personal data shall be [...] collected for specified, explicit and legitimate purposes and not further processed in a manner that is incompatible with those purposes'' (Art. 5.1, referred to as \textit{purpose limitation})
    \item \textbf{Revertible}: ``The data subject shall have the right to withdraw his or her consent at any time. [...] It shall be as easy to withdraw as to give consent.'' (Art. 7.3)
    \item \textbf{Unburdensome}: ``[...] a declaration of consent pre-formulated by the controller should be provided in an intelligible and easily accessible form, using clear and plain language and it should not contain unfair terms'' (Recital 42)
\end{itemize}

However, the seeming inability of consent to curb the problems associated with surveillance capitalism has caused others to question its presentation as a catch-all solution; the rise of business models driven by targeted advertising means that not all parties involved in the consent process desire its successful operation, leading to `consent theatre' as companies craft user experiences designed to increase the likelihood that users click accept~\cite{fassl2021stop, zuboff2019age}. This view is supported by studies showing that the design of consent interfaces significantly affects acceptance rates (i.e., consent is not freely given) and that almost all written material provided to users making consent decisions is too complex to be easily understood (i.e., consent is not informed)~\cite{utz2019uninformed, luger2013consent, ma2022prospective}. Suggested responses target every aspect of the consent process, ranging from technical solutions such as reducing the number of decisions by applying pre-defned policies~\cite{fassl2021stop} and making privacy policies more accessible~\cite{luger2013consent}, through to re-framing consent as a means of informing users, rather than just a disclosure exercise~\cite{luger2013informed} and integrating practices from BDSM communities such as periodically checking in to see if interactions are meeting users' expectations~\cite{strengers2021what}. Others have explored allowing users to delegate consent decisions to third parties, although found that around 50\% of users still wished to make decisions themselves~\cite{nissen2019should}.

Previous work on VFC for Alexa has highlighted four key issues with its present implementation in relation to the principles of consent outlined above~\cite{seymour2022can}: the time pressure introduced by the eight second response (consent is not freely given), the limited amount of information that can be usable delivered via speech (consent is not informed), the interface used different flows to give and withdraw consent (consent is not revertible) and that trusted speech originating from consent mechanisms is indistinguishable from untrusted speech from third-party skills. The use of voice for permissions---rather than devices like smartphones which already include an authentication layer---also raises questions about whether the \textit{correct person} is giving consent. The introduction of voice `profiles' for individual users represents an attempt to address this,\footnote{E.g., \url{https://us.amazon.com/gp/help/customer/display.html?nodeId=GYCXKY2AB2QWZT2X} (accessed 18/11/2022)} but prior work has reported high error rates with VA voice profiles trained to recognise a particular person~\cite{huang2020amazon}.

But from a legal perspective, consent is not the only reason that organisations can collect people's data: the GDPR describes six legal justifications (bases) that can be used for data collection, including performance of a contract and legitimate interests (e.g., fraud prevention) in addition to the consent of the data subject. While less of a problem in interpersonal interactions, a key aspect of consent in HCI and Ubicomp is that the user knows who they are giving consent \textit{to}. The GDPR uses the language of data subjects, controllers, and processors to define the relationship of different parties to information that is collected, the definitions of which are given in Table~\ref{tab:gdprDefs}. The GDPR also places additional requirements on the processing of `special category' data, processing of which requires ``explicit consent [...] for one or more specified purposes'' that may not be met by VFC (Art. 9, examples of special categories include ethnicity, health, and political opinions).

\begin{table}
    \centering
    \caption{Definitions from Article 4 of the GDPR~\cite{GDPR16}. Text has been edited for clarity (e.g. by removing enumerated examples).}
    \begin{tabularx}{1.0\columnwidth}{l|X}
        \toprule
        Term & Definition \\ \hline
        Personal Data & Any information relating to an identified or identifiable natural person (`data subject'). \\
        Data Processing & Any operation or set of operations which is performed on personal data or on sets of personal data. \\
        Data Controller & The natural or legal person which determines the purposes and means of the processing of personal data. \\
        Data Processor & A natural or legal person which processes personal data on behalf of the controller. \\
        Consent & Any freely given, specific, informed and unambiguous indication of the data subject’s wishes by which they, by a statement or by a clear affirmative action, signifies agreement to the processing of personal data relating to them. \\
        \bottomrule
    \end{tabularx}
    \label{tab:gdprDefs}
\end{table}

\subsection{Permissions and Privacy Labels}\label{sec:priv-background}
The technical (and often conflated) counterpart to consent, permissions are the most common way of facilitating access to data on smart and mobile devices. But the complexity of connected devices makes effectively communicating relevant information a difficult task; research into the efficacy of smartphone permissions on platforms like Android and iOS has historically shown problems with attention and understanding~\cite{felt2012android}. Users also have more nuanced responses than just declining permissions---such as choosing different apps or minimising app installations---and these choices can be traced back to more fundamental attitudes and intentions~\cite{alsoubai2022permission}.

Complementing work on informed consent, research into privacy notices and labels that succinctly communicate important information to users has emerged in response to poor comprehension of permissions. General work identifies the \textit{timing} of these notices, whether they come via the same \textit{channel} as data collection, their \textit{modality}, and whether they integrate user \textit{controls} as key factors when designing privacy labels~\cite{schaub2017designing}. The nature of privacy means that these criteria are contextual and will often be intertwined, such as how the timing of location sharing notifications affects both users' ability to control disclosures as well as their ability to enact their privacy preferences~\cite{patil2014reflection}. In general, privacy notices should accompany meaningful choices in order to avoid the type of `dejected acceptance' often seen in studies on privacy perceptions~\cite{schaub2015design, schaub2017designing, shklovski2014leakiness}.

Standardisation is also an important factor in promoting accuracy and speed when interpreting privacy notices, highlighting the importance of their implementation at the platform level~\cite{kelley2010standardizing}. On smartphone platforms many of these findings have been adopted by app stores, including privacy labels and nudging developers to remove permissions not requested by functionally similar apps~\cite{peddiniti2019reducing, li2022understanding}. Some platforms also allow users to restrict the granting of permissions to when the app is foregrounded or grant one-time permissions which expire when the app is closed.\footnote{\url{https://developer.android.com/training/permissions/requesting} (accessed 08/12/2022).} While there is no widely implemented equivalent for smart home devices, research soliciting expert opinions has identified how devices are updated and whether they uses default passwords as important information to be included on potential labels~\cite{emami2020ask}.

Comparing smartphone app permissions to the voice assistant equivalents, the former control access to device functionality (e.g., microphone or storage) while skill permissions explicitly control access to personal data (e.g., name or address). This creates two major differences: (1) granting app permissions will not always expose personal data (e.g., the ability to pair with Bluetooth devices, or location access whilst in airplane mode) making them a constraint on what an app can \textit{do}, but many VA permissions directly relate to personal data (e.g. the address that an Alexa device is located at); and (2) personal data gained through app permissions could be utilised solely on-device (e.g. using a microphone for local transcription), but the architecture of contemporary voice assistants means that personal data must be sent to third-party skills via the internet. This subtle shift in the framing of permissions moves voice assistant consent mechanisms closer conceptually to legal consent processes than the traditional access control mechanisms that preceded mobile permissions on personal computers.

\section{Methods}\label{sec:methods}
To answer the research questions given above we conducted a Delphi study with experts from academia, industry, and the regulatory and policy sector. While the specifics of the Delphi methodology vary, the core protocol involves a group of ``[anonymous] experts who are invited to assess and comment on different statements or questions related to a specific research topic''~\cite{beiderbeck2021preparing}. Participants comment on their and others' responses and are given the option of revising their opinions in a process that continues over a number of rounds. Due to the exploratory nature of the research questions and complex problems under consideration, our study protocol was based on the \textit{policy Delphi}, which focuses on exploring options and supporting evidence rather than just reaching consensus~\cite{turoff1975delphi}. Crucially, as well as identifying areas of consensus between participants we also wanted to capture areas of disagreement that might warrant additional investigation in future work. Full details of the survey and analysis are provided as supplemental material and archived at \url{https://osf.io/4vu67}.

We chose this approach because the topic represents a complex design space: best practices for voice interfaces are still emerging and contemporary platforms have niche technical limitations; at the same time, the GDPR imposes strict legal constraints on how organisations must manage consent but does not specifically address verbal consent. Our RQs specified experts as they already understand the broader technical and regulatory landscape and have the experience necessary to evaluate and narrow-down a broad initial selection of requirements for VA verbal consent. This provides a solid foundation of requirements for follow-up work with end-users, developing those requirements further to meet their specific needs.

\subsection{Participant Recruitment}
Eight participants were recruited following recommended best practices of having an heterogeneous group of experts~\cite{boulkedid2011using} by reaching out to academics, practitioners, and policy/regulatory experts via previous connections and publicly available contact details. Those contacted were also asked to pass details of the study on to their own contacts with appropriate expertise (snowball sampling). To avoid deanonymising our participants, we describe their combined expertise: our expert panel drew on many decades of experience (1) conducting research on voice assistants and online privacy and security at globally leading universities; (2) at a senior level in global organisations focused on voice assistants and online privacy and security; and (3) policy and regulation of technology around online privacy and security. To give an indication of their current focus participants were also asked to self-describe their current roles, which are shown in Table~\ref{tab:participants}, but many participants had represented several of these perspectives over the course of their careers. While none of the participants were practising lawyers, they had considerable experience with relevant privacy law from the three perspectives above and the manuscript was reviewed by a senior legal scholar prior to submission. While facilitators of previous Delphi studies in HCI have themselves participated in the ratings and discussions~\cite{jones2015telling}, we elected not to do so in order to prevent undue influence on the results given that we had created the initial study statements. Participants remained anonymous during the study and were given the option not to be directly quoted in publications. All parts of the study were approved by our institution's ethics review board.

\begin{table}
    \centering
    \caption{Participant's self-described job roles.}
    \begin{tabularx}{1.0\columnwidth}{l|X}
    \toprule
    P1 & PhD Student specialising in Privacy and Digital Market Power \\
    P2 & Security Business Development (security, compliance, identity, and privacy capabilities) \\
    P3 & Researcher specialising in usable security and privacy \\
    P4 & Researcher specialising in voice assistant security and privacy \\
    P5 & Associate Professor specialising in data protection, machine learning, and the regulation of technology \\
    P6 & Senior Data Privacy Consultant \\
    P7 & Assistant Professor specialising in security, privacy, and HCI \\
    P8 & Policy Officer \\
    \bottomrule
    \end{tabularx}
    \label{tab:participants}
\end{table}

\subsection{Statement Curation}
As is common in many Delphi studies we curated a list of statements in advance in order to streamline the study and reduce the number of data collection rounds required~\cite{vogel2019delphi}. From the research questions we identified: (1) data protection legislation that governs the use of consent as a legal basis for collecting and processing data~\cite{GDPR16}; (2) HCI and Ubicomp research that explores issues around consent and privacy notices~\cite{luger2013informed, strengers2021what, fassl2021stop, cheng2020dangerous, schaub2017designing}; and (3) research ethics guidance on informed consent~\cite{oxford0000researchethics}. Based on standard Delphi methodology, these sources were used to create 41 candidate statements relating to verbal consent for voice assistants, with each item formulated as a modal statement. A full list of statements is given in Table~\ref{tab:allstatements} in the Appendix. These statements were not anchored to a specific voice assistant, with participants asked to consider voice assistants in general when approaching the statements.

\begin{itemize}
    \item For regulation, statements represented individual stipulations. E.g., ``\textit{Where personal data relating to a data subject are collected from the data subject, the controller shall, at the time when personal data are obtained, provide the data subject with all of the following information: [...] the period for which the personal data will be stored, or if that is not possible, the criteria used to determine that period}'' from GDPR Art. 13~\cite{GDPR16} became ``\textit{Verbal consent should say how long data will be stored/the conditions used to determine how long to store it}''.
    \item For research papers, statements were created for each recommendation. E.g., ``\textit{The ability to review/withdraw data relates to the ability for users to review and withdraw their consent, and therefore their data, at any point during or after their interaction with a system. By allowing for multiple `choice' points, such measures can support users' voluntary choices.}'' from \cite{luger2013informed} became ``\textit{Users should be prompted to renew consent granted via voice at regular intervals}''.
    \item For research ethics guidelines, statements reflected each prompt recommended for inclusion in an oral consent script. E.g. ``\textit{How identifiable you will be: [Explain how easy it will be for them to be identified from any publications or other research outputs.]}'' from ~\cite{oxford0000researchethics} became ``\textit{Verbal consent should include how identifiable users will be from the data collected}''.
\end{itemize}

\subsection{Delphi Rounds and Analysis}
We implemented the main rounds of the study as an online survey using Qualtrics. The order of the statements was randomised and participants were asked to rate each statement for its \textbf{relevance} to the VA verbal consent process, it's \textbf{actionability} in the design and development process, and how \textbf{usable} it would be for end users. Answers were given as three Likert items coded from 1 (Strongly disagree) to 5 (Strongly agree). Participants were also encouraged to leave comments describing the reasoning behind their answers. This is common best practice in Delphi studies as it enables investigators to better understand the views of participants and allows participants to identify new statements that should be added to subsequent rounds~\cite{boulkedid2011using}.

Quantitative analysis of participant ratings was conducted using Python. Responses for relevance, actionability, and usability were evaluated in terms of agreement between participants, and for statements with high agreement, the central tendency of participants' ratings. The interquartile range (IQR) of participants' ratings was used as a measure of agreement for individual items, which accounts for the ordinal nature of Likert items and is generally considered robust when analysing Delphi studies~\cite{VonDerGracht2012}. Median rating values were used to identify the central tendencies of participant responses. This is a robust measure of central tendency for ordinal data, and use of the mean in this context can be problematic as it implies that the `distance' between responses is uniform across the item and can be skewed by outlier responses~\cite{VonDerGracht2012}.

For the second round our focus was on generating discussion on areas of disagreement from round one. Because presenting ratings and comments from round one alongside the original statements greatly increased the amount of time required to respond to the survey, in round two we narrowed the focus to statements with the lowest per-item agreement to keep the total time required similar to round one. No new statements were generated from the four feedback comments we received from participants but two free text questions were added asking about potential privacy-related questions users might ask a VA (for Statement 24) and general data-sharing rules that might be set (for Statement 34). These comments also led to the clarification of legal bases as described in Section~\ref{sec:roleofconsent}. Participants were presented with a bar chart of results from round one and the anonymised comments from other participants relating to the statement, as recommended when reporting the results of previous rounds back to participants~\cite{boulkedid2011using} (see Figure~\ref{fig:round-two-sample}). Participants were then given the option to reconsider their ratings based on the results from other participants and asked to leave a new comment explaining their reasoning. We re-ran the quantitative analysis after the conclusion of round two. For free text comments, the Delphi method is itself the method of analysis through soliciting, aggregating, and refining expert opinions~\cite{turoff1975delphi}. As such, we report the findings for this part of the study by summarising the conversations between experts.

In round three we followed up with participants individually instead of continuing the format of rounds one and two, given that our objective was to identify broad areas of (dis)agreement and very few items remained above the IQR disagreement threshold after round two. We asked participants specific questions relating to topics they had raised in the comments in order to further develop and understand their arguments. Of particular interest were topics or concepts that significantly influenced or would be influenced by adoption of the survey statements (such as joint controllership, as detailed below). As these questions supplement the discussion in round two, we report them in Section~\ref{sec:disagreements-open-questions}.

\section{Round One Results: Agreement and Understanding the Role of Consent}
\subsection{Areas of Agreement}\label{sec:agreement}
In the first round, there were 7 statements that received very high ratings for relevance, actionability, and usability ($\tilde{x} \geq 4.5$, this was the lowest threshold that selected under 25\% of the statements) with good agreement ($IQR \leq 2.0$). Participants generally did not comment on these statements, feeling that they did not need to justify their choices in these cases. These statements were:

\begin{itemize}
    \item Verbal consent should say how users can withdraw their consent (S12)
    \item Verbal consent should be clearly distinct from interactions with third-party skills (e.g. spoken in a different voice, S15)
    \item Granting of verbal consent should require a clear affirmative statement (S16)
    \item Verbal consent should come with voice commands that revoke a skill's access to personal data (S30)
    \item Verbal consent should say if data will be used to track them on the voice assistant platform or elsewhere on the internet (S32)
    \item Platforms should require that all skills using verbal consent publish a privacy policy (S35)
    \item Platforms should regularly verify that links to privacy policies remain valid (S36)
\end{itemize}

As might be expected, these statements are not controversial. We see that they align closely with the principles of consent outlined by \citeauthor{im2021yes}~\cite{im2021yes}: consent should be voluntary (S15, S16), informed (S32, S35, S36), revertible (S12), and unburdensome (S30). Many are also legally required in some form, such as the publishing of privacy policies by entities controlling personal data --- although the lack of verification of those policies as in S36 in the voice assistant ecosystem has been the subject of recent research~\cite{guo2020skillexplorer,young2022skilldetective,edu2021skillvet, edu2022measuring}. A clear affirmative action is a legal requirement for consent under the GDPR (Art. 4), and Article 7 specifies that ``it shall be as easy to withdraw as to give consent''~\cite{GDPR16}. While what measures would satisfy this in the case of VAs is a matter of interpretation, the provision of voice commands arguably achieves this. There is, however, no requirement to inform users about \textit{how} they can withdraw their consent beyond the existence of their rights as data subjects (Art. 13). Relating the results to prior work, we note that S35 \& S36 align with~\cite{guo2020skillexplorer,young2022skilldetective,edu2021skillvet, edu2022measuring}, and that S12, S15, S30, and S36 align with the changes to verbal consent suggested in~\cite{seymour2022can} as well as the potential for multi-agent voice assistants that clearly differentiate between agents through voice~\cite{zargham2021multi} (particularly S15). These statements also hint at potential implementations that would mostly require superficial changes to current implementations (e.g., changing the speech prompts used at various stages of the verbal consent process)---we discuss these further in Section~\ref{sec:discussion}.

\subsection{The Nature of Consent as a Legal Requirement and Ethical Best Practice}\label{sec:roleofconsent}
In other areas, participants were unable to agree on the statements and began to discuss the reasoning behind their different positions. A key discussion emerged in the first round of the study around the legal role that consent plays in verbal dialogues. The first point made by several participants was that using consent as the legal basis for data collection imposed a number of strict requirements. Therefore, if the goal was to obtain consent only using speech, then the process would be unusably verbose and many of the statements in the survey would be legal obligations: ``\textit{The kinds of consents you’re asking for are required by law. There isn't really any question about it. They have to be made actionable and usable, or the data can’t be gathered legally}'' (P6). Participants also highlighted that the use of other legal bases for data collection could be problematic, such as the presentation of consent-style options to users when consent was not itself the legal basis: ``\textit{If legitimate interests is the basis, then it's not consent, so it's misleading to talk about consent in this context}'' (P5).

This made it apparent that from a legal perspective the most likely way that verbal consent could be implemented for VAs would be a written privacy policy underpinning a verbal consent dialogue. It would then be a matter of interpretation as to how sufficient attention could be drawn to written documents in order to ensure compliance. The issue of other legal bases is also an important consideration given that legal grounds such as \emph{contract} are likely to be used for voice assistant skills (where collected data is used to fulfil a legal contract with the user, such as the taxi example in Table~\ref{tab:sampleVFC}). Guidance from the UK data protection regulator suggests that ``If you would still process the personal data without consent, asking for consent is misleading and inherently unfair'' and that ``If you make consent a precondition of a service, it is unlikely to be the most appropriate lawful basis''.\footnote{\url{https://ico.org.uk/for-organisations/guide-to-data-protection/guide-to-the-general-data-protection-regulation-gdpr/lawful-basis-for-processing/consent/} (accessed 18/11/2022)} This is not just a problem for verbal consent---the predominant model for permissions on smartphones and other platforms similarly presents every request as if it were a consent request. We return to this point in Section~\ref{sec:discussion}.

We had left the legal basis of data collection deliberately ambiguous in the first round in order to prevent priming participants, hoping to surface discussions such as this and observe any assumptions made. While this was successful in stimulating discussion in round one, feedback from participants suggested that this ambiguity was preventing them from fully engaging with some of the statements. We therefore asked participants to assume that consent was the legal basis for data collection underpinned by a written privacy policy when answering questions for round two. This discussion around the legal role played by consent highlighted the other important role that it played, which was conceptualised by participants as an \emph{ethical best practice}. First identified from cases where consent was not the legal basis used for processing, there were instances where participants believed that it could be beneficial to e.g. let users set general rules for data sharing even though this was not legally required or might not have legally compliant implementations.

\subsection{Joint Controllerships}
When considering what information should be present in verbal consent dialogues there was an unexpected discussion on the potential for joint controllerships, where more than one organisation effectively has control over the use of personal data. The GDPR contains a provision for controllers who ``jointly determine the purposes and means of processing'', mandating that they ``in a transparent manner determine their respective responsibilities [...] in particular as regards the exercising of the rights of the data subject'' as well as the information that controllers are required to provide to data subjects (Art. 26). However, when asked about VAs verbally stating the identity of the data controller(s), participants suggested that the situation regarding joint controllers was still developing: ``\textit{Identifying the data controller is a necessary, but not trivial exercise. Recent jurisprudence highlights that there is often a joint controllership situation, particularly in mobile apps and websites.}'' (P1), ``\textit{Maybe not so actionable if the controller(s)/processor(s) haven't actually decided who is what and whether they are joint controllers}'' (P5). P8 suggested that the identity of the controller(s) was a good example of information that was more useful when provided in written form (e.g., to facilitate making a complaint).

\begin{figure*}
    \centering
    \caption{Medians for relevance, actionability, and usability ratings given by participants. Where a statement was re-rated in the second round of the study, a line links the median of the re-ratings with the corresponding median from round one.}
    \includegraphics[width=1.0\textwidth]{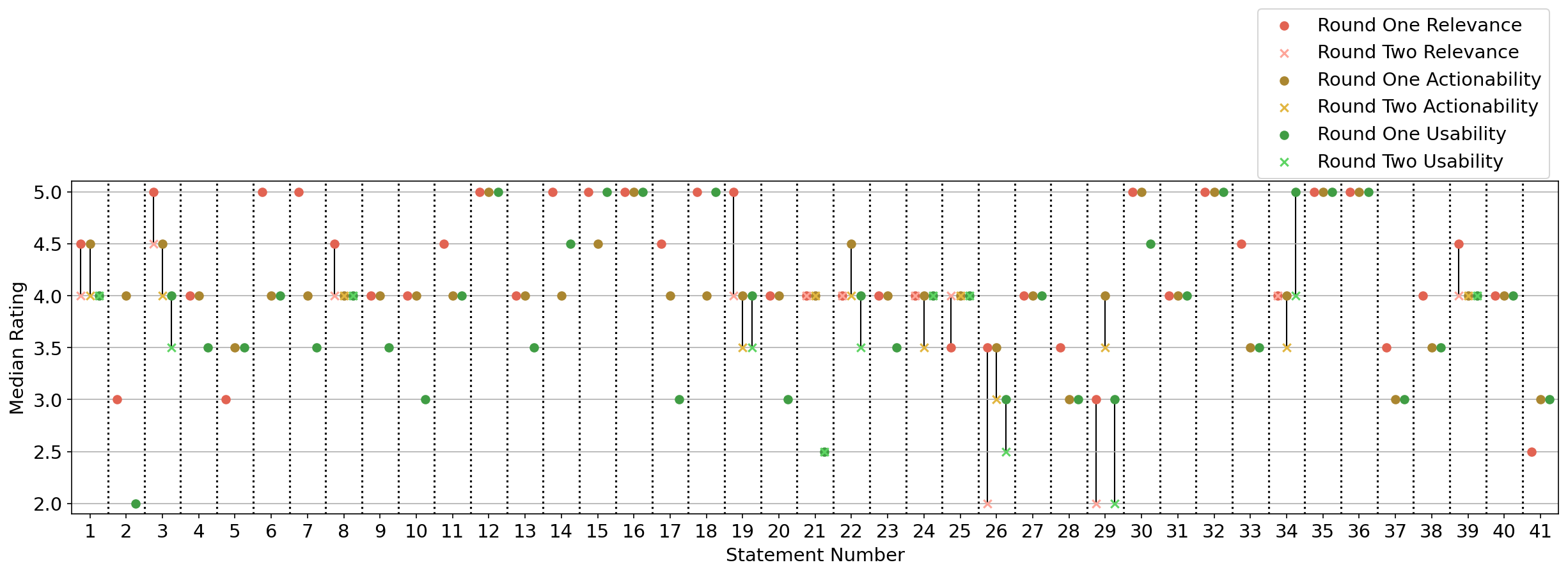}
    \label{fig:ratings-mean}
\end{figure*}

\begin{figure*}
    \centering
    \caption{Interquartile ranges for relevance, actionability, and usability ratings given by participants. Where a statement was re-rated in the second round of the study, a line links the IQR of the re-ratings with the corresponding IQR from round one.}
    \includegraphics[width=1.0\textwidth]{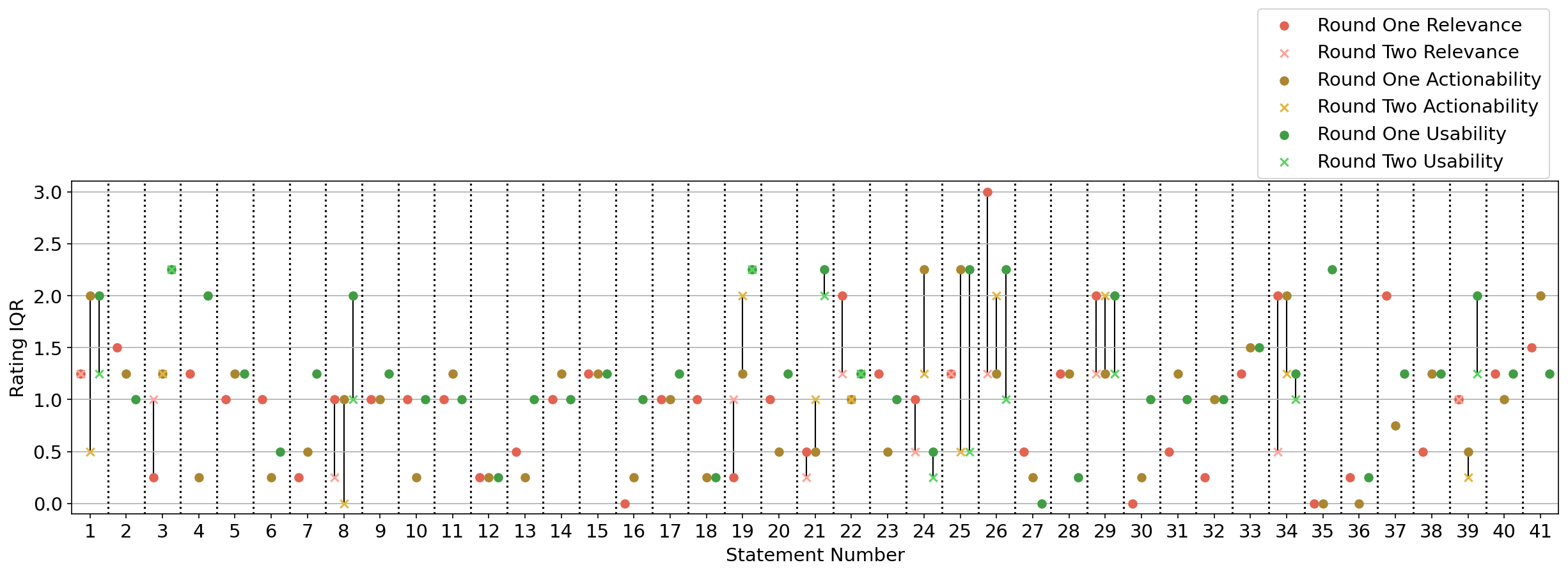}
    \label{fig:ratings-iqr}
\end{figure*}

\section{Round Two Results: Disagreements and Open Questions}\label{sec:disagreements-open-questions}
After the conclusion of the second round we calculated median responses and interquartile ranges for each of the Delphi statements in order to answer RQ1, which are shown in Figures \ref{fig:ratings-mean} and \ref{fig:ratings-iqr} respectively. Considering RQ2, we turned to the disagreements between participants. When selecting statements to include in round two, we focused on those items with higher disagreement ($IQR \geq 2.00$), which returned 21 items across 16 statements. To avoid separating individual items from their wider context we included each of relevance, actionability, and usability items for a statement so long as ratings for at least one were above the disagreement threshold. As per the unexpected discussion around the role of consent reported in Section~\ref{sec:roleofconsent}, in the second round we asked participants to assume that consent was the legal basis for data collection and that this was backed by a written privacy policy covering legal obligations. We therefore excluded four of the selected statements that contradicted this position. For the 42 of the 48 included Likert items, the additional review and discussion in round two either lowered or did not alter the level of disagreement between participants.

\subsection{Introducing Nuance to Consent Decisions}
Several statements pertained to platforms offering users consent decisions that went beyond a simple binary or framed the VA verbal consent process as a negotiation or conversation. Participants were quick to point out the problems with similar attempts in the past ``\textit{The cookie experience has shown how difficult it can be to introduce requirements to choose levels of access, it leads to companies using long lists to discourage users from opting out.}'' (P2) and that---particularly in the context of VA skills---the principle of data minimisation in Article 5 of the GDPR limited the data that developers could collect ``\textit{Permissions should only be asked for if they’re needed for the defined purpose the skill is intended to complete.}'' (P6). This led to a view that consent was treated ``\textit{mostly a compliance exercise and not about giving users control}'' (P1).

Despite the fact that some participants suggested that purpose limitation was not effective in practice, the statement asking if VA platforms should nudge users towards skills that used fewer permissions (effectively enforcing purpose limitation themselves, S26) was not well received, with participants questioning its relevance and suggesting that it distracted from the problem of developers requesting too much data. In fact, nudging was seen as antithetical to the concept of informed consent: ``\textit{In some ways it goes against the idea of consent as it bypasses the user's capacity as a rational agent}'' (P5), ``\textit{Nudging and obtaining consent are two different concepts; I could nudge the user towards giving consent to dangerous skills}'' (P7). P1 suggested that the focus on permissions distracted from the ``\textit{contextuality and individuality in privacy [...] a lot of aspects of data collection aren't really easy to anticipate ahead of time. While legal requirements might prescribe such an ex ante exercise, these can be quite burdensome for users and not give them real control}'', suggesting that giving users more general system-wide defaults for certain behaviours such as tracking would provide more genuine control (explored further in Section~\ref{sec:opportunities} below).

\subsection{What to Include in Consent Dialogues}
A major discussion across the Delphi rounds centred on the information that should be included in VA verbal consent dialogues (S1-14, 19, 20). Succinctly communicating all of the information typically found in a privacy policy is generally infeasible for \textit{any} interface, and the need for parsimony is even greater for VAs given the lower bandwidth of speech compared to graphical interfaces. Overall, participants believed that while much of this information was relevant to the consent process, it may not be usable (especially in the legal language used by documents like privacy policies).

Participants could not agree about including the name of the data controller in consent dialogues. It was highlighted that this knowledge is essential when exercising information rights, but in practice it could be difficult to maintain this in a way that was usable (as mentioned in Section~\ref{sec:roleofconsent}): ``\textit{I still ask what that means to a user who may not know? It might be more relevant to say who will have access to the data}'' (P2). In addition to including the organisation(s) that would gain access, P8 suggested signposting the part of the privacy policy containing the controller's contact details to allow for the exercise of rights set out by Chapter 3 of the GDPR (including the right to access and erasure). There were similarly mixed responses to the inclusion of processing purposes, data types, and retention periods, with P8 suggesting that in most cases where special category data is not collected it would be sufficient to direct users to the privacy policy for more information. While current implementations of VA permissions are relatively concrete (e.g., the user's address), it is conceivable that in the future this may not be the case, and participants were concerned they could become less easy to understand. Again, while all participants agreed that processing purposes were relevant, no consensus was reached around whether or how they should be included verbally, with the main discussion around the benefit to users of always hearing purposes.

\subsection{Opportunities Created Through Speech}\label{sec:opportunities}
Beyond ways to improve the verbal consent process in line with conventional equivalents, there was also discussion about how to best leverage the opportunities that might arise through the use of speech and the general architectures of current VAs. The first of these was the use of metaphors to explain aspects of the consent process. While metaphors can be used textually, their delivery via speech could potentially be much more natural and engaging. P6 was initially opposed to their use on the basis that they were unnecessary. This comment generated a lot of push back in the second round: ``\textit{I think metaphors are actually highly relevant. Privacy might be straightforward for experts like us, but hardly for the average user}'' (P1), ``\textit{I disagree with the statement that the concepts of data privacy are straightforward. Although the principle might be the practical, implications of sharing data isn't so providing someway for users to understand what this could mean for their privacy is important. I'd need to understand the plans to see if metaphors is the best approach for this}'' (P2). This sense of cautious optimism was reflected in P6's own reflection in round two, where they explained that ``\textit{my worry is that [metaphors will] be used as a way to obscure the truth, rather than reveal it}'', acknowledging that ``\textit{it is important that privacy policies are couched in terms appropriate to the user, so if there's a way to do it really well that improves understanding, then go for it}'' (P6). Along similar lines P8 suggested that metaphors may be appropriate when presented alongside proper legal terminology, and P7 that future research could develop and validate the effectiveness of potential metaphors.

Participants also discussed whether it made sense to utilise the conversational nature of VAs to let users ask questions about data collection, effectively turning the consent process into a dialogue (S24). Participants were tentatively positive about the potential to engage users with their rights in a more natural way, but echoing prior work on privacy bots~\cite{harkous2016pribots} voiced major concerns about the engineering challenge of creating an assistant that could accurately answer questions without introducing further uncertainty. While several participants gave examples of open-ended questioning, signposting for a more limited set of questions was suggested by P7, and P8 gave examples of how this could be tied to specific aspects of legal consent (e.g., purposes, bases, and identities of data recipients).

Other statements in this category made use of the casual nature of conversational interfaces to suggest ways that consent might be treated more as an ongoing process. With regards to having users renew/revisit their consent decisions on subsequent uses of a skill (S22 \& S29), ratings and comments on these statements suggested that participants did not think either would be very relevant or usable for similar reasons: ``\textit{consent will be renewed at regular intervals --> leading to customer fatigue / habituation / lack of attention}'' (P7), ``\textit{while consent is important, there's no point in annoying the user beyond what's necessary. Asking once should suffice. So, no reason to change my mind}'' (P6).

As was expected given the above responses, participants were much more enthusiastic about Statement 34, which proposed allowing users to set general rules about when they wanted to share data with skills in an attempt to reduce the overall number of consent decisions. While this was considered usable and ``\textit{would be a lot against consent fatigue}'' (P1), participants were concerned about actionability and the legal implications of platforms making automated data sharing decisions on users' behalf: ``\textit{Consent has to be linked to purpose. Different skills will have different purposes, so you can't just collect blanket consent}'' (P6) with P8 agreeing that any potential rules would have to be specific and informed. When asked for examples about the kinds of rules it would be beneficial for users to set, participants gave a combination of positive and negative formulations e.g., ``\textit{End-users might be asked when first setting up the device whether they are fine with apps tracking them, or accessing other pieces of information (and then never again). I also think end-users should only be asked about aspects that they can understand and can control}'' (P1). The requirement that potential rules were understandable was echoed by others, who also stressed possible impacts on transparency, suggesting that ``\textit{there should be a quick notification / confirmation of sharing}'' when rules were triggered (P2).

\section{Discussion: Towards A New Model of Consent for Voice Assistants}\label{sec:discussion}
We now discuss the areas where consensus emerged around general principles and specific practices to provide recommendations for regulators, first-party platforms, and third-party developers. Beyond these initial changes to VFC implementations, and to fully answer RQ3, we also take a longer-term view of consent through conversational interfaces. Based on the responses from our participants and prior work in the field, we make six recommendations that we believe can more closely align verbal consent with privacy regulation and ethical best practices. Together these recommendations cover changes to platform architectures and data protection regulations. We also discuss other contexts to which they might prove valuable, such as smartphones.

\subsection{Requirements with Broad Support}
Section~\ref{sec:agreement} presented a number of requirements for VA verbal consent that promote legal and ethical standards without requiring significant technical or regulatory changes. Our expert participants deemed each of these requirements highly relevant to the consent process, easy to action by VA platforms, and usable by voice assistant users.

The consent dialogue itself should be articulated in a voice clearly distinct from the one used by third-party skills so as to demark trusted platform-originated speech from untrusted dialogue from third-party developers. It should also require a clear affirmative statement from the user, describe how consent can later be withdrawn, and mention whether the skill will track the user on the voice assistant platform or elsewhere on the internet. VA platforms should include voice commands that revoke consent, require skills using personal data to publish a privacy policy, and regularly check that these are adequate and remain up to date.

\subsection{Matching Mechanisms with Legal Bases}
A repeated point of contention for participants was the conflation of permissions mechanisms with legal consent. At present, every request for information via VFC asks the user if they give third-party skills permission to access their personal data, even if this is not the legal basis under which that data is collected. This is misleading and creates situations where users might potentially `refuse consent' only to have that data legally collected by other means, and as such conflates consent with data protection in its entirety. The way that VAs present data collection should align with the legal bases and purposes under which that data is collected.

\subsubsection*{Recommendation 1: Distinguish between data collected under different legal bases}
This primarily entails not presenting data collected under grounds such as legitimate interest or contract as a consent decision, such as through the use of verbal statements requiring explicit approval from the user. It is, however, important that users are aware that a skill is using their personal data regardless of the legal basis used and still have the option not to use a skill---for example the Ride Hailer skill in Table~\ref{tab:sampleVFC} would likely use contract as the legal basis for data collection and processing, as the user's address and mobile number are required in order to provide the taxi service. To this end, we propose including a short voice snippet when the skill is first used that either points users to a card sent to their device or briefly outlines the data that will be accessed. This approach has the additional benefit of reducing the number of consent decisions that need to be made---moving them to be \textit{non-blocking} in the language of~\cite{schaub2015design}. Such an approach could be considered problematic as it requires developers to fully understand the legal intricacies around how they collect and process personal data, but the current architecture of VA platforms already makes developers legally responsible for data collected. Data collected on the basis of consent would continue to utilise mechanisms similar to VFC. To promote transparency, companion apps should allow users to see recent uses of personal data (similar to Apple's iOS Privacy Report feature). For example:

\begin{quote}
    \textit{This skill accesses your personal data. See the card in your Alexa app for more information.}
\end{quote}

\begin{quote}
    \textit{This skill will access your [address] in order to [fulfil a contract with you]. Say `Alexa, stop' to close the skill.}
\end{quote}

\subsubsection*{Recommendation 2: Make clear upfront when data sharing is a precondition of service}
The above would also allow for a meaningful upfront distinction to be made around whether sharing data with a skill is optional or required in the context of an interaction by differentiating between data collected on the basis of consent and under other bases. At present, users have little indication whether refusing to share data with a skill will immediately end the interaction or have it continue in a modified way. Developer guidelines say that skills should `gracefully' handle the refusal of consent via the API (i.e. they should not crash), but guidance from data protection regulators such as the UK data protection regulator state that making consent a precondition of service is unlikely to be appropriate.\footnote{\url{https://ico.org.uk/for-organisations/guide-to-data-protection/guide-to-the-general-data-protection-regulation-gdpr/lawful-basis-for-processing/consent/} (accessed 18/11/2022)}

This distinction is important because agreeing to data processing essential for functionality is not the same as consenting to sharing for e.g., targeted advertising; by only using VFC for data gathered using the legal basis of consent and using other mechanisms where data sharing is a precondition of service, the implications of refusing to share would be clearer to users. This would also make it easier to certify skills by clarifying that gracefully handling refusal of consent should mean that a skill can still be used. This could also reinforce the principle of purpose limitation by dissuading skills from requesting data that they do not need in order to complete their function. 

\subsection{Reducing the Number of Consent Decisions}
There is a general narrative of `consent fatigue' on smartphones and the web whereby the sheer frequency of complex privacy decisions with little immediate impact drives people to accept privacy policies with little thought. With conversational interfaces this burden is increased when users are forced to use companion apps to make consent decisions. The introduction of VFC may increase usability, but does not change the number of decisions that people need to make. Even if people need to make comparatively few consent decisions through voice interfaces compared to elsewhere online, it is important to minimise the overall frequency of consent decision making; participants were clear that consent fatigue was one of the major problems when creating verbal consent experiences, and that reducing the frequency of decisions was very important.

\subsubsection*{Recommendation 3: Utilise approaches such as social norms to reduce the number of consent decisions users must make}
One of the promising proposals from the study was to allow users to specify general rules about data sharing when they configure a platform or device for the first time. While this idea is not new, prior work on similar access controls in smart homes has shown that they are typically underutilised by users, with \citeauthor{zeng2019understanding} finding that social norms were more important for users when negotiating access to shared devices~\cite{zeng2019understanding}. To this end, we include them here as one of many possible means of reducing the number of consent decisions needing to be made by users. Recent work on exploring and capturing these norms offers inspiration for solutions that are based around these expectations of how VAs should behave more generally and whether specific instances of data sharing would be appropriate~\cite{abdi2021privacy}. Norms could be mined in advance from a representative set of people based on contextual integrity~\cite{nissenbaum2004privacy} and used as defaults. Users could also be given the opportunity at first use (or later) to choose from clusters of norms that usually occur together, have their norms inferred by answering indicative questions about example scenarios, or have them automatically refined over time~\cite{malkin2022runtime,zhan2022model}. These norms would then be applied to requests for data sharing by skills with users notified as appropriate (perhaps through a notice similar to those above). When none of the selected norms apply, then the standard mechanisms would be used (this includes VFC but could also extend to data sharing where consent is not the legal basis being used), which could also feedback to keep adapting to the user. Being able to identify norms used to make data sharing decisions in advance would both dramatically reduce the number of choices that need to be made in real time, as well as helping to build trust and alleviate the `creepiness' of voice assistants due to their perceived violations of social norms around listening and use of data~\cite{lau2018alexa}. Although the most usable way to represent and convey norms to users is an active area of research, potential examples of norms include:

\begin{quote}
    \textit{Datatype=Email → Unacceptable} (from~\cite{abdi2021privacy}).
\end{quote}

\begin{quote}
    \textit{Datatype=Location} and \textit{Category=Transport → Acceptable} (based on Table~\ref{tab:sampleVFC}).
\end{quote}

\subsubsection*{Recommendation 4: Give third parties greater scope to consent on behalf of users}
A major obstacle to the implementation of norm-based consent is the GDPR itself, which requires that consent must be ``specific, informed and unambiguous'' (Recital 32). A blanket opt-in consent such as that granted by the second norm above is not specific or informed, as it could relate to an infinite number of potential purposes and recipients. Recital 42 states that ``\textit{for consent to be informed, the data subject should be aware at least of the identity of the controller and the purposes of the processing for which the personal data are intended}'', but neither the identity of the data controllers nor the purposes for which they may wish to process the data can be known when giving consent in advance in this way. Making purposes standardised across a platform and machine readable to facilitate this would be technically challenging and logistically infeasible.

Despite this, the GDPR does not forbid a third-party from giving consent on an individual's behalf provided that they have the authority to do so and that the above requirements for consent are met. Tools that allow people to make blanket decisions in advance already exist, such as the Global Privacy Control specification,\footnote{\url{https://globalprivacycontrol.org} (accessed 18/11/2022)} the EU Interactive Digital Advertising Alliance `Your Online Choices' tool,\footnote{\url{https://www.youronlinechoices.com} (accessed 18/11/2022)} and the US Digital Advertising Alliance's `Your Ad Choices' tool.\footnote{\url{https://youradchoices.com} (accessed 18/11/2022)} The key difference is that these tools process opt-outs rather than opt-ins (i.e., they are not concerned with the \textit{granting} of consent). Another potential issue is that such a system would be controlled by first-party platforms (e.g., Amazon and Google) who have a considerable conflict of interest with regards to people using their products; the EU and US tools above were created by industry self-regulation bodies which have similar conflicts of interest, but cannot opt users into data sharing. An alternative arrangement would be to have tools run by consumer advocacy groups or non-profits to foster trust, although the locked-down nature of current VA platforms would make this difficult. Prior work suggests that alongside competence, perceived intentions and moral integrity of third parties are important factors affecting peoples' trust in delegated consent decisions~\cite{nissen2019should}. It is unfortunate that the GDPR does not easily allow for this kind of automated decision making to alleviate consent fatigue, but the vested interests of platforms in collecting data would make this a difficult area to effectively regulate.

\subsection{Promoting Interface Symmetry and Policy Enforcement}
\subsubsection*{Recommendation 5: Provide voice commands to withdraw consent}
The GDPR makes it clear that it must be equally easy to withdraw consent as to give it. Nielsen's usability heuristics similarly describe how users ``\textit{need a clearly marked `emergency exit' to leave [an] unwanted action without having to go through an extended process}'',\footnote{\url{https://www.nngroup.com/articles/ten-usability-heuristics} (accessed 18/11/2022)} and a subsequent adaptation for voice assistants similarly states that ``\textit{users often choose system functions by mistake and will need an option to effortlessly leave the unwanted state without having to go through an extended dialogue}''~\cite{langevin2021heuristic}. When consent can be given via voice, people should therefore be able to easily and intuitively withdraw it in the same way, as there are likely to be cases where consent is given in error. In practice this means providing voice commands that revoke a skill's access to user data and signposting their existence when users give verbal consent (S30). In terms of more technical implementation, Alexa skills operate on a model of \textit{intents}. These allow developers to script responses to certain actions a user might want to take, but they also ensure that certain commands will work with any skill; the `stop' intent, for example, will always end an interaction with a skill. Providing a similar intent for consent revocation would allow for this to happen at any point during interaction, and give users confidence as this is enforced by the platform rather than relying on third-party developers.

\subsubsection*{Recommendation 6: Hold platforms accountable for hosted skills}
The final recommendation sits between regulation and platform policy, and involves deliberately considering the role that VA platforms and developers play in relation to data protection. By design, the GDPR governs the relationship between an individual data subject and the legal person who controls their data, but this does not reflect how VA platforms operate in practice. While third-party developers are indeed the ones in receipt of personal data, the way that they gain consent, maintain it, and even access the data that they `control' is governed (or at least mediated) by the VA platform on which they operate. In the case of Alexa, they must use Amazon's consent API and accept Amazon's placement of their privacy policy and other details on the Alexa store. Developers are even forbidden from storing the data they control, instead having to access it through the provided customer profile API before each use. Yet the responsibility for adhering to legal and ethical data protection obligations is pushed entirely onto developers. This echoes wider concerns over content moderation on online platforms and any response needs to be carefully considered accordingly.

A clear way that this issue manifests is with privacy policies. While Amazon mandates that skills using personal data publish a privacy policy and checks this during the certification process, research over a number of years has shown that in many cases these policies are not checked \textit{after} certification and are now missing or defective~\cite{guo2020skillexplorer,young2022skilldetective,edu2021skillvet, edu2022measuring}. While developers are and should be responsible for publishing privacy policies that comply with the law, platforms should also have a degree of responsibility for ensuring that the content they host is compliant. As reported in~\cite{edu2022measuring}, a responsible disclosure to Amazon reporting 246 skills with problematic data practices led to decisive action for many skills, yet around 40\% of them still had issues a year later.

\subsection{Applicability to Other Devices and Platforms}
While this work focuses on voice assistants there are a number of other contexts where the recommendations above could offer improvement. As described in Section~\ref{sec:priv-background}, smartphone apps use a similar permissions model (it is highly likely that the current Alexa store and permissions system was modelled on the equivalent Android and iOS features), meaning that recommendations one to three apply to varying extents for these platforms.

Beginning with recommendation one, all `dangerous' permissions that can be requested by Android apps are presented under the guise of consent (e.g., ``Allow TikTok to access your contacts? <Yes> <No>''), whilst other permissions are granted silently by the operating system. According to recommendation one, a framing of consent should only be used when the user can meaningfully use the app after declining. In line with recommendation two, permissions sought via `consent' when apps will not function without them effectively makes sharing data a precondition of service and means that other lawful bases are likely to be more appropriate, but current implementations cannot differentiate between them. Android, for example, does automatically allow requests for permissions not flagged as `dangerous', but these are more aligned with whether the functionality provides access to data rather than why it is being used.

Recommendation three has an obvious application to smartphone apps---allowing users to specify general rules around data sharing in advance could significantly address consent fatigue by reducing the number of data sharing decisions that users have to make. This is especially the case when research suggests that far more smartphone apps request permissions associated with personal data than VA skills~\cite{edu2021skillvet, kollnig2022iphones}. It is unclear at present how automated consent decisions based on norms could be legally implemented on smartphones (see recommendation four), but the greater extensibility of smartphone operating systems means that permission decisions could conceivably be handled by user-specified applications.

While recommendations five and six address concerns specific to the two major voice assistant platforms, similar concerns have been expressed for mobile platforms over the moderation of and data collection by third-party apps, although these discussions tend to be more mature than for voice assistants (e.g.~\cite{shklovski2014leakiness}). Since 2012, when Google announced that ``open systems win'' in relation to an Android app store that did not have a certification process for apps, there have been a host of automated and manual checks introduced for apps before they can be made publicly available (such as Android's `Play Protect'.\footnote{\url{https://developers.google.com/android/play-protect} (accessed 18/11/2022)}).

\section{Limitations and Future Work}
As mentioned in Section~\ref{sec:methods}, we chose to adopt a broad perspective in our research questions and study design which led to the identification of issues applicable across a wide range of use cases for verbal consent. Follow-up work will be required for the different contexts in which VA verbal consent is used, as each will have a different set of unique issues in addition to the high level concerns identified in the study. Solutions to these must also be adapted with and for the values of individual communities. While the European position adopted in relation to regulation potentially limits the applicability of the findings to other jurisdictions, as noted in Section~\ref{sec:introduction} several other countries have adopted regulations considered compatible with the GDPR.

In terms of future work there were some important topics surfaced, such as the power asymmetries between users, developers, and first-party platforms that did not generate enough support or discussion to fully explore within the remit of the study. This was more often the case for statements unaligned with regulation, which understandably tended to be opposed by participants from a compliance background. While this perspective was an invaluable part of the study, it created a higher bar for support for some statements not backed by the weight of the law. This is not to say that compliant statements were above scrutiny; examples such as the push back on metaphors and lack of support for some statements derived from the GDPR demonstrate how participants were willing to argue both ways for what they believed. We hope that others can continue to explore these additional aspects of VA verbal consent. For instance, in the case of power asymmetries future research could consider the perspectives of those disadvantaged by such asymmetries. This will, in turn, help shape the regulatory landscape of the future.

\section{Conclusion}
As voice assistant platforms mature and strive to improve the user experience it seems inevitable that additional functionality will move from companion apps into the conversation. The results of the study show how taking this approach with consent requires skilful navigation of legal and ethical requirements that need to be balanced against user experience considerations. Through a Delphi study with subject experts, we make six recommendations that use the novelties and constraints of speech to propose new ways of approaching VA verbal consent. At the same time, our participant discussions open up a range of new questions for future work that the Delphi method is less equipped to investigate. These include the considerable power wielded by platforms in pushing the legal responsibility for data protection onto developers whilst simultaneously setting the terms on which this must be done, and how this is in turn creates norms around data collection and use. Above all, we think it is an opportune time to extend and diversify the ways we conceptualise consent for voice assistants and other conversational technologies, especially as they seem likely to become an ever-more common mode of interaction.

\begin{acks}
This work was undertaken as part of the Secure AI Assistants project through Engineering and Physical Sciences Research Council grant EP/T026723/1. We would also like to thank Perry Keller and Hana Kopecka for reviewing drafts of the manuscript.
\end{acks}

\balance

\bibliographystyle{ACM-Reference-Format}
\bibliography{main}

\appendix
\section{Appendix}

\begin{figure*}
    \centering
    \caption{Sample question presented to participants in round two.}
    \includegraphics[width=0.7\textwidth]{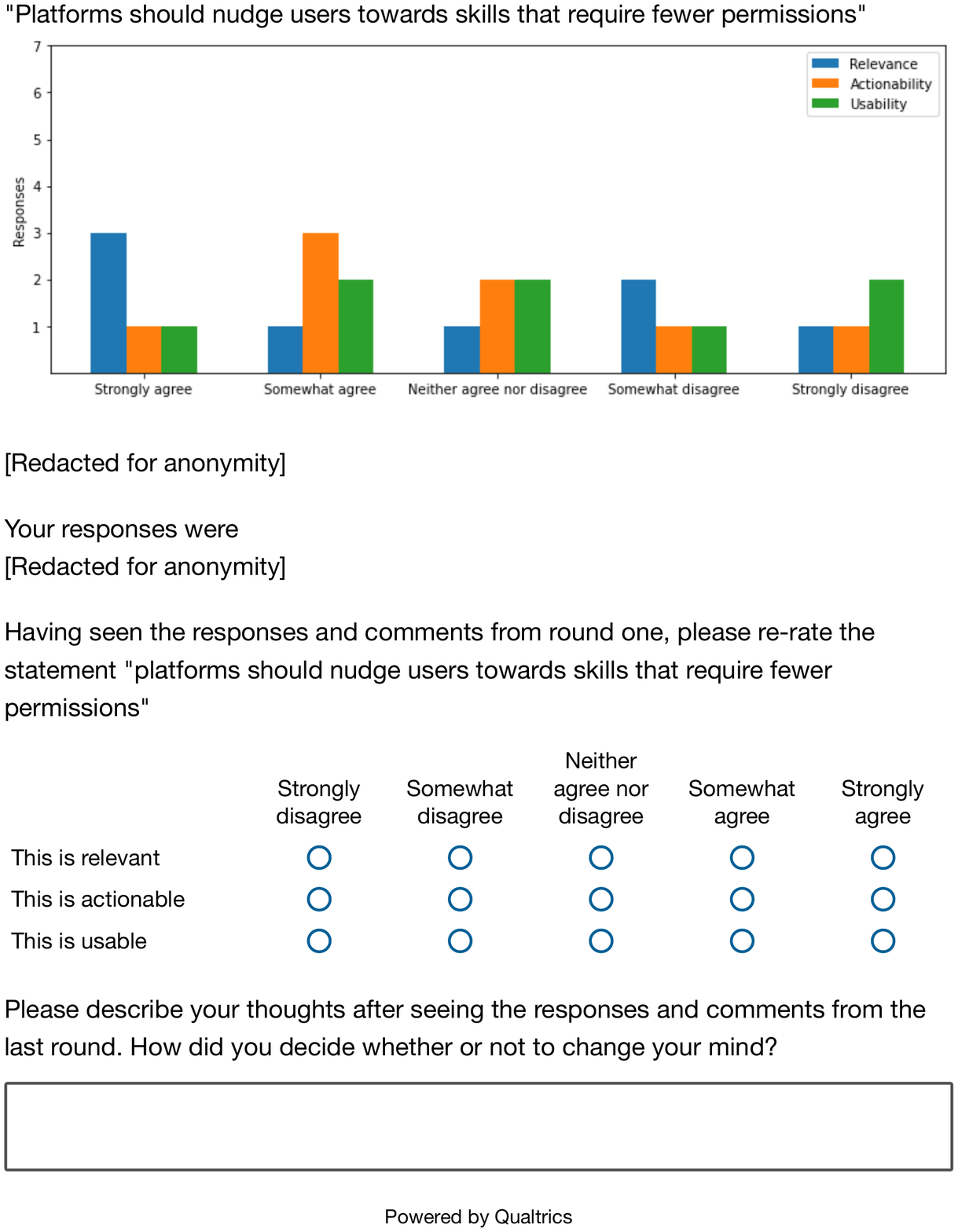}
    \label{fig:round-two-sample}
\end{figure*}

\begin{table*}
    \centering
    \caption{Complete list of statements used in the study. Statements included in round two are shown in \textbf{bold}.}
    \begin{tabularx}{1.0\textwidth}{r|X}
    \toprule
    \textbf{S1} & Verbal consent should say the identity of the data controller \\
    S2 & Verbal consent should say the contact details of the data controller \\
    \textbf{S3} & Verbal consent should say the purpose(s) data will be used for \\
    S4 & Verbal consent should say the legal basis on which the data is processed \\
    S5 & Where this is legitimate interests, verbal consent should say what those legitimate interests are \\
    S6 & Verbal consent should say the identities of other parties the data will be shared with \\
    S7 & Verbal consent should mention if the data will be transferred to a third country \\
    \textbf{S8} & Verbal consent should say how long the data will be stored (or the conditions used to determine how long to store it) \\
    S9 & Verbal consent should remind users of their right of access \\
    S10 & Verbal consent should remind users of their right of erasure \\
    S11 & Verbal consent should say how users can complain about data processing \\
    S12 & Verbal consent should say how users can withdraw their consent \\
    S13 & Verbal consent should mention if automated decision-making will be used \\
    S14 & Verbal consent should say whether the data controller will process the data for any other purposes \\
    S15 & Verbal consent should be clearly distinct from interactions with third party skills (e.g. spoken in a different voice) \\
    S16 & Granting of verbal consent should require a clear affirmative statement \\
    S17 & Where processing has multiple purposes, consent should be given for all of them \\
    S18 & Granting verbal consent should not be unnecessarily disruptive to the current interaction \\
    \textbf{S19} & Verbal consent should say the types of personal information being shared \\
    S20 & Verbal consent should say how identifiable users will be from the data collected \\
    \textbf{S21} & Verbal affirmations of consent should be sought for each type of data shared with a skill (rather than one affirmation for all data) \\
    \textbf{S22} & Users should be prompted to renew consent granted via voice at regular intervals \\
    S23 & Verbal consent should be renewed when new functionality is added to a skill \\
    \textbf{S24} & Verbal consent should be a two-way process, with users able to ask questions about data collection \\
    \textbf{S25} & Verbal consent should make use of metaphors where appropriate to help people understand how their data will be used \\
    \textbf{S26} & Platforms should nudge users towards skills that require fewer permissions \\
    S27 & Companion apps should allow users to visualise how their data is shared with skills \\
    S28 & Verbal consent should offer intermediate options beyond a binary yes or no \\
    \textbf{S29} & Verbal consent should prompt users to confirm their consent the second time they use a skill  \\
    S30 & Verbal consent should come with voice commands that revoke a skill's access to personal data \\
    S31 & Platforms should enforce standards around how skills ask for verbal consent \\
    S32 & Verbal consent should say if data will be used to track them on the voice assistant platform or elsewhere on the internet \\
    S33 & Verbal consent should frame consent as an opportunity to negotiate what kinds of data a skill recieves \\
    \textbf{S34} & Verbal consent should allow people to specify general rules for sharing their data to reduce the number of consent decisions \\
    S35 & Platforms should require that all skills using verbal consent publish a privacy policy \\
    S36 & Platforms should regularly verify that links to these privacy policies remain valid \\
    S37 & Verbal consent should primarily inform and empower users (e.g. rather than fulfil legal obligations) \\
    S38 & Verbal consent should be triggered when data is requested/needed (e.g. rather than on skill launch) \\
    \textbf{S39} & Platforms should highlight unusual permissions requests when gathering verbal consent \\
    S40 & There should be guidance from a regulator (e.g. the ICO) on how platforms should utilise verbal consent \\
    S41 & Consent should not be the legal ground used for data collection by skills (e.g. rather contract or legitimate interests) \\
    \bottomrule
    \end{tabularx}
    \label{tab:allstatements}
\end{table*}

\end{document}